\documentclass[reprint,aps,prd,preprintnumbers,nofootinbib]{revtex4-1}

\usepackage[sumlimits,intlimits,namelimits]{amsmath}
\usepackage{amssymb}
\usepackage[english]{babel}
\usepackage{bm}
\usepackage{epstopdf}
\usepackage{graphicx}
\usepackage{slashed}
\usepackage[usenames,dvipsnames]{xcolor}

\begin{document}

\title{$\bm{1/m}$ Corrections for Orbitally Excited Heavy Mesons and the $\bm{1/2}$ -- $\bm{3/2}$ Puzzle}
\author{Rebecca Klein}
\email{klein@physik.uni-siegen.de}
\author{Thomas Mannel}
\email{mannel@physik.uni-siegen.de}
\author{Farnoush Shahriaran}
\email{shahriaran@physik.uni-siegen.de}
\author{Danny van Dyk}
\email{vandyk@physik.uni-siegen.de}
\affiliation{Theoretische Physik 1, Naturwissenschaftlich-Technische Fakult\"at,
Universit\"at Siegen, Walter-Flex-Stra\ss{}e 3, D-57068 Siegen, Germany}

\preprint{SI-HEP-2015-04, QFET-2015-05}

\begin{abstract}
We re-investigate the effects of the $1/m_c$ corrections on the spectrum
of the lowest orbitally excited $D$-meson states.
We argue that one should expect the
$1/m_c$ corrections to induce a significant  mixing between the two lowest lying $1^+$ states.
We discuss the implications of this mixing and compute its effect on the semileptonic decays
$B \to D^{**} \ell \bar{\nu}$ and the strong $D^{**}$ decays.
\end{abstract}

\maketitle

\section{Introduction}

The spectroscopy of excited hadrons containing a heavy quark is determined to a large extend by the
fact that the spin of the heavy quark decouples from the light degrees of freedom
\cite{Isgur:1991wq}. To this end, the rotations of the heavy quark spin become a symmetry that is
not present for light hadrons. As a consequence, all heavy hadrons (with a single heavy quark) fall
into spin-symmetry doublets, the members of which are related by a rotation of the heavy quark spin.

For the mesonic ground states,  the spin symmetry doublets consist of the $0^-$ pseudo scalar meson
and the $1^-$ vector meson, such as $(D,D^*)$ and $(B,B^*)$. For the mesons with one additional unit
of angular momentum of the light degrees of freedom two further spin symmetry doublets appear.
These correspond to a total angular momentum $j = 1/2$ and $j = 3/2$ of the light degrees of
freedom, respectively.  Coupling it to the heavy quark spin and taking into account the factor
$(-1)^{\ell + 1}$ for the parity, we end up with a doublet with the quantum numbers $(0^+,1^+)$ for
$j = 1/2$ and one with $(1^+,2^+)$ for $j = 3/2$.

The same systematics continues for higher orbital excitations, i.e.\ for  arbitrary $\ell$
\cite{Isgur:1991wq}. The doublets which emerge have the quantum numbers
\begin{equation*}
    ((\ell-1)^{(-1)^{\ell + 1}},\ell^{(-1)^{\ell + 1}}) \quad \text{for}\quad j = \ell-1/2\,,
\end{equation*}
and
\begin{equation*}
    (\ell^{(-1)^{\ell + 1}},(\ell+1)^{(-1)^{\ell + 1}}) \quad \text{for}\quad j = \ell+1/2\,.
\end{equation*}

The decoupling of the heavy quark also has interesting consequences for the strong decays of excited
heavy hadrons, since these are then governed by the light degrees of freedom only. In this way one
may predict the partial waves involved in the strong decays and thus one may predict a specific
pattern of angular-momentum suppressions in the strong decays \cite{Isgur:1991wq,Lu:1991px}.

In the heavy mass limit the members of the doublets are mass-degenerate, and the splittings between
the multiplets are mass independent.  Nevertheless, these splittings are of the order of
$\Lambda_{\rm QCD}$, since they are related to excitations of the light degrees of freedom.

The $1/m_c$ corrections couple the heavy quark spin to the light degrees of freedom, leading also to
a splitting within the doublets. However, in particular for the $D$ mesons, the splitting between
the multiplets and the ($1/m$ induced) splittings within each of the multiplets are numerically of
the same order. Thus a significant mixing of the two $\ell^{(-1)^{\ell + 1}}$ states belonging to
the two different multiplets must be expected.

Previous analyses of the $1/m_c$ corrections also noticed the possibility of this mixing
\cite{Lu:1991px,Kilian:1992hq,Falk:1995th,Leibovich:1997em,Bernlochner:2012bc}.  However, some of
the analyses assumed that this effect is small, based on the observation that the two $1^+$ states
should have different widths, since the strong decays of the $1^+$ state with $j = 3/2$ correspond
to $p$-wave transitions. Since the data indicate that the observed widths indeed follow this
pattern, the mixing was assumed to be small.

In the present paper we re-consider this effect and try to estimate the size of the corresponding
mixing angle, which can be extracted from the data on the basis of a few model assumptions. Based on
this, we extract a mixing angle that is larger than what has been discussed before, and
consider the implications in particular for semileptonic $B$ decays into orbitally excited $D$
mesons.

\section{Orbitally Excited States and the Effect of Mixing}

The orbitally excited states fall into two spin-symmetry doublets, classified by their angular
momentum of the light degrees of freedom $j$. We shall use the notation
\begin{equation}
\label{D}
\begin{split}
    \left(\begin{matrix}
    | D(0^+) \rangle \\
    | D(1^+) \rangle
    \end{matrix} \right)  \quad & \text{with} \quad j = 1/2\,,\\
\left(\begin{matrix}
| D^*(1^+) \rangle \\
| D^*(2^+) \rangle
\end{matrix} \right) \quad & \text{with} \quad j = 3/2\,.
\end{split}
\end{equation}
In the limit $m_c \to \infty$ the members within each of the spin symmetry doublet are degenerate:
\begin{equation}
\begin{aligned}
        M(D(0^+)) = M(D(1^+)) & = m_c + \bar\Lambda_{1/2}\,, \\
    M(D^*(1^+)) = M(D^*(2^+)) & =  m_c + \bar\Lambda_{3/2}\,,
\end{aligned}
\end{equation}
where $\bar\Lambda_{j} $ is the binding energy of the mesons in the limit $m_c \to \infty$.

Note that the splitting $ \bar\Lambda_{3/2} - \bar\Lambda_{1/2}$ between the two doublets does not
scale with the heavy quark mass. However, it is related to the binding of the light quark within the
chromoelectric field of the heavy quark and hence it is of the order of $\Lambda_{\rm QCD}$. In
fact, for the $D$ mesons, the current data yield a value of about 20 MeV for this splitting.

Power corrections of order $1/m_c$ are induced by the kinetic and the chromomagnetic operator,
leading to a Hamiltonian density of the form
\begin{equation} \label{1om}
    {\cal H}_{1/m}
    = \frac{1}{2 m_c} \bar{c} (i D_\perp)^2 c + \frac{g_s}{2 m_c}   \bar{c} (\vec{\sigma}\cdot
    \vec{B}) c\,.
\end{equation}
In particular, the second term couples the heavy quark spin to the light degrees of freedom, which
breaks the spin symmetry; however,  the angular momentum of the light degrees of freedom $j$ is
still a good quantum number.

We are going to consider the effects of ${\cal H}_{1/m}$ on the two spin symmetry doublets shown in
eq.~(\ref{D}). First of all, the kinetic energy contribution only leads to a shift of the masses,
which we shall absorb into the values of the masses
\begin{equation} \label{kin}
    M_{j} = m_c + \bar\Lambda_j + \frac{1}{2 m_c} \mu_\pi^2 (j)\,,
\end{equation}
where $\mu_\pi^2 (j)$  is the kinetic-energy parameter for the orbitally excited meson.  The second
term in eq.~(\ref{1om}), however, leads to the effects, which we shall discuss in some detail. It is
related to the chromomagnetic field induced by the light degrees of freedom at the location of the
heavy quark spin.  Clearly not much is known about this from first principles, so one has to make
some assumptions here to arrive at quantitative estimates.

The chromomagnetic field at the location of the heavy quark is generated by the angular momenta of
the light degrees of freedom, which is the orbital angular momentum $\vec{L}$ and the spin of the
light quark $\vec{s}$, the sum of which constitutes the angular momentum of the light degrees of
freedom $\vec{J}$.  The complete angular momentum $\vec{K}$ is
\begin{equation}
\vec{K} = \vec{J} + \vec{\sigma} = \vec{L}+\vec{s}+\vec{\sigma}\,.
\end{equation}
The key assumption we shall make is that the orbital angular momentum and the light quark spin have
different gyro-chromomagnetic factors $\alpha'$ and $\beta'$, such that the chromomagnetic field
seen by the heavy quark is not proportional to the total angular momentum $J$ of the light degrees
of freedom:
\begin{equation} \label{Blight}
\vec{B} \sim \alpha'  \vec{L} + \beta ' \vec{s} = \alpha \vec{J} + \beta \vec{s}\,.
\end{equation}
To this end, the second term in eq.~(\ref{1om}) can be written as
\begin{equation} \label{Hoom}
H_{1/m} = \int d^3 \vec{x}\,  \frac{g_s}{2 m_c}   \bar{c} (\vec{\sigma}\cdot \vec{B}) c
=  P_1  (\vec{J} \cdot \vec{\sigma} ) + P_2  (\vec{s} \cdot \vec{\sigma} )\,,
\end{equation}
where the operators $P_1$ and $P_2$ act only on the radial wave functions of the light degrees of freedom.

We note that the $| D^*(2^+) \rangle$ and the $| D(0^+) \rangle$  states  are eigenstates of the
Hamiltonian, even once the  above $1/m_c$ corrections are included
\begin{align}
    H  | D^*(2^+) \rangle & = \left(M_{3/2} + \frac{3}{4} g + \frac{1}{4} g' \right)  | D^*(2^+) \rangle\,, \\
    H  | D(0^+) \rangle   & = \left(M_{1/2} - \frac{3}{4} g + \frac{1}{4} g' \right)  | D(0^+) \rangle\,,
\end{align}
where the mass values  $M_{1/2}$ and $M_{3/2}$
are the masses defined in eq.~(\ref{kin}).
Furthermore, the constants $g$ and $g'$ are obtained from the radial wave
functions, i.e. the matrix elements $g \sim \langle P_1 \rangle$ and $g' \sim \langle P_2 \rangle$;
here we assume for simplicity that $g$ and $g'$ are identical for
the two doublets. The relevant coefficients are obtained from the spin wave
functions discussed in the appendix.

For the two $1^+$ states, the second term of eq.~(\ref{Hoom}) induces a mixing, since we have
\begin{gather}
\begin{aligned}
    H | D(1^+) \rangle
    & = \left(M_{1/2} + \frac{1}{4} g - \frac{1}{12} g' \right)  | D(1^+) \rangle\\
    & \quad + \frac{\sqrt{2}}{3} g'  \, | D^*(1^+) \rangle\,,
\end{aligned}
\\
\begin{aligned}
    H | D^* (1^+) \rangle
    & = \left(M_{3/2} - \frac{5}{4} g - \frac{5}{12} g' \right)  | D^*(1^+) \rangle\\
    & \quad + \frac{\sqrt{2}}{3} g' \,  | D(1^+) \rangle\,,
\end{aligned}
\end{gather}
where we have -- once more -- assumed that the relevant matrix elements of $P_1$ and $P_2$ are identified with
the parameters  $g$ and $g'$. Again, the relevant coefficients in front of $g$ and $g'$ follow from
the spin wave functions given in the appendix.

\begin{figure}[h]
    \centering
    \includegraphics[width=0.35\textwidth]{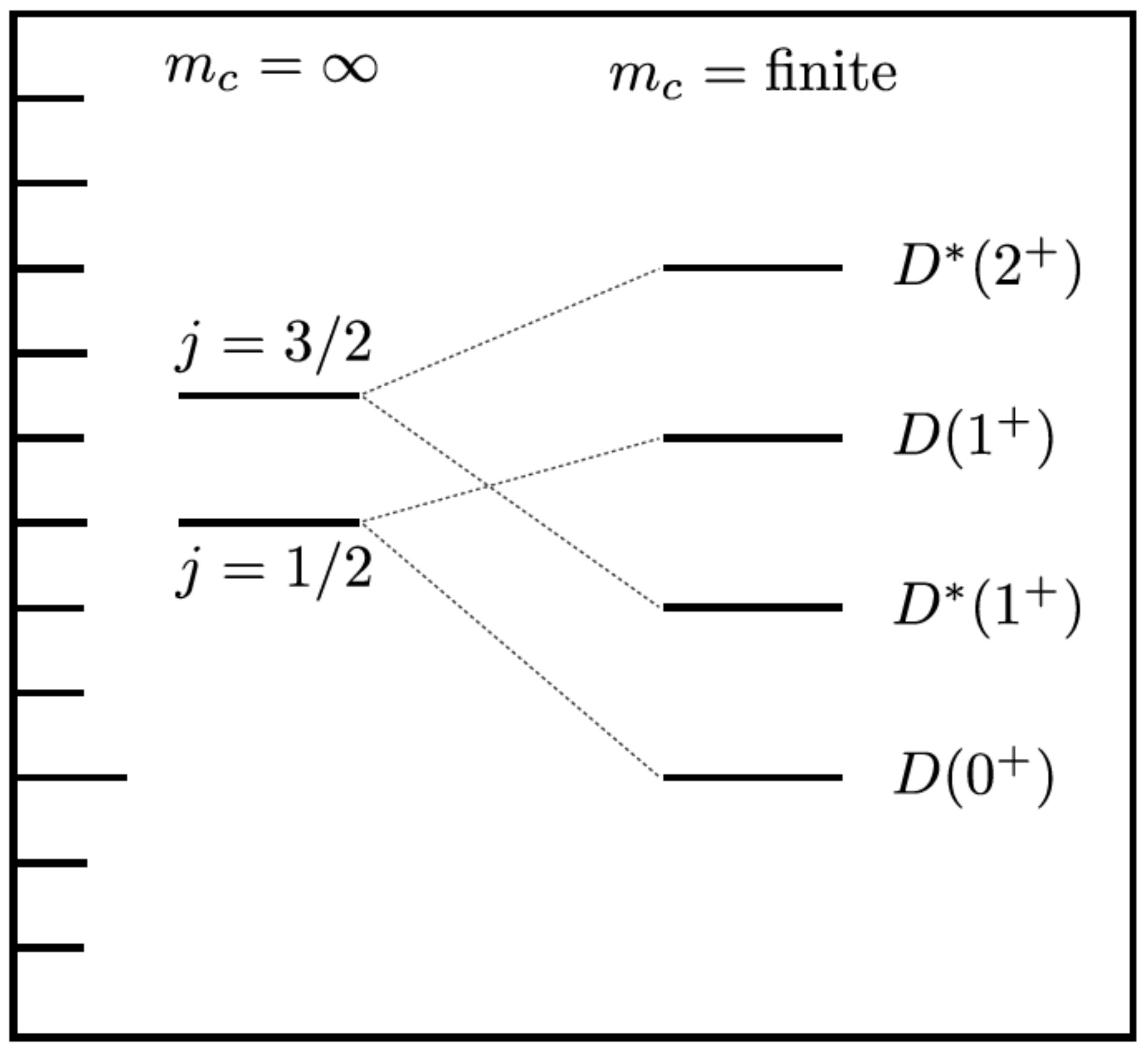}
    \caption{Schematic of the mass hierarchy for the orbitally excited $D$ mesons.}
    \label{splitt}
\end{figure}

Clearly this analysis is drastically simplified, but an improvement would need the reference to some
model for the binding dynamics of the orbitally excited mesons. Nevertheless, from this simplified
analysis we can infer some information on the mixing angle from the data. First of all -- looking at
the data summarized in table~\ref{tab1} and schematically shown in figure~\ref{splitt} -- we note
that the broader $1^+$ state has a larger mass than the narrower $1^+$ state. From heavy-quark
symmetries we infer that the broader states decay through an $S$ wave transition which is possible
only for the $j = 1/2$ states. The $j=3/2$ states can decay (strongly) only through a $D$ wave
transition which is suppressed by angular momentum, rendering these states narrow
\cite{Isgur:1991wq}. For this reason we make the assignments shown in  table~\ref{tab1}.

\begin{table}
\begin{center}
\renewcommand{\arraystretch}{1.3}
\begin{tabular}{|c|c|c|c|c|}
    \hline
    State      & Mass [MeV]         & Width [MeV]       &  $j$  \\
    \hline
    \hline
    $D(0^+)$   & $2318   \pm  29$   & $267   \pm  40$   & $1/2$ \\
    $D (1^+)$  & $2427   \pm  40$   & $384   \pm 120$   & $1/2$ \\
    \hline
    $D^*(1^+)$ & $2421.4 \pm   0.6$ & $ 27.4 \pm   2.5$ & $3/2$ \\
    $D^*(2^+)$ & $2462.6 \pm   0.6$ & $ 49   \pm   1.3$ & $3/2$ \\
    \hline
\end{tabular}
\end{center}
\caption{Averages of the existing mass and width measurements for the $D^{**}$ states
according to the PDG \cite{Agashe:2014kda}. Our choice of the $j$ assignment follows from
total decay widths.}
\label{tab1}
\end{table}

\begin{figure}[t]
    \centering
    \includegraphics[width=0.45\textwidth]{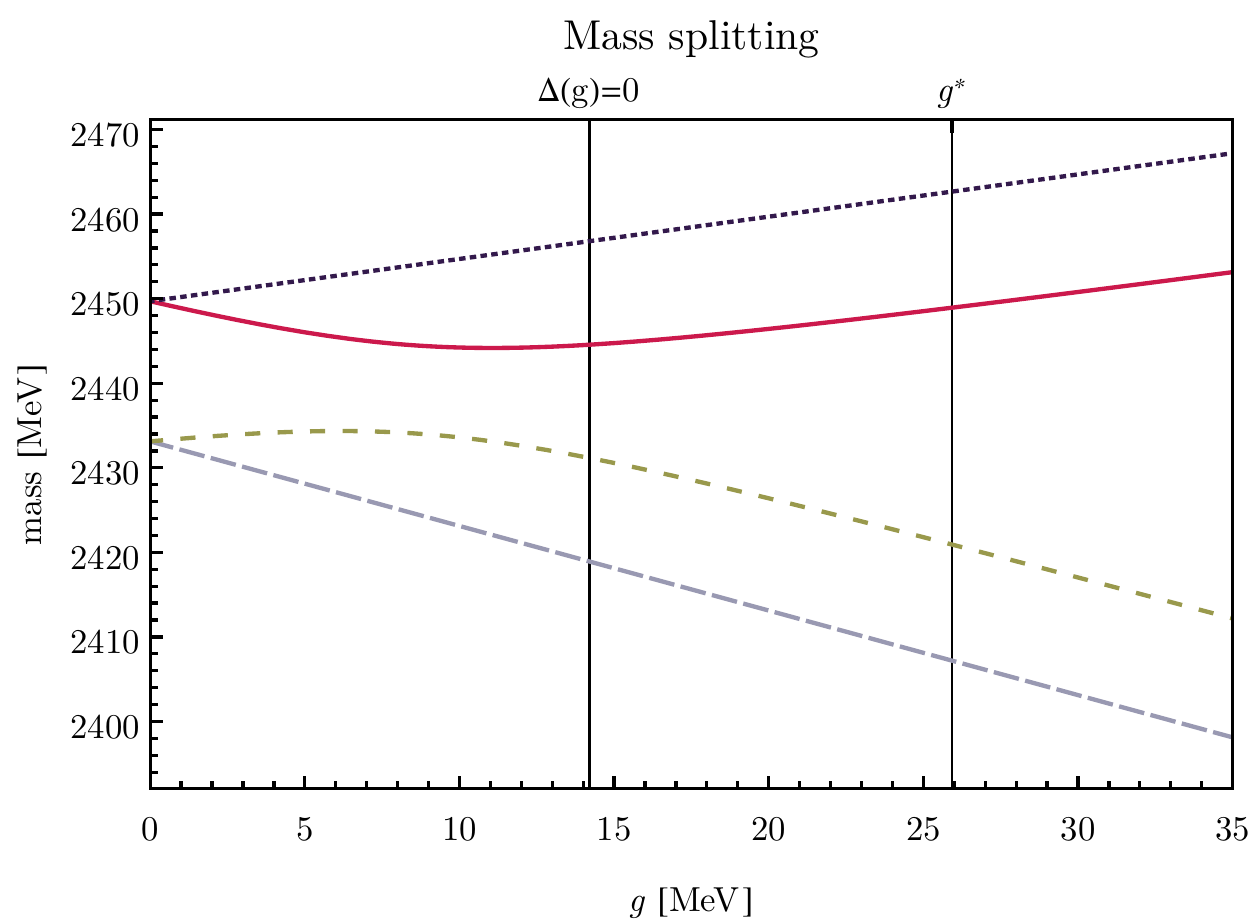}
    \caption{The orbitally excited states in the $j=1/2$ and the $j=3/2$ doublets are degenerate in the
          infinite mass limit $m_c=\infty$, corresponding to $g = g' = 0$.
          For a finite mass $m_c$ we find a splitting of
          the states. We plot the masses of the orbitally excited states
          $D(0^+)$ (light-blue, long-dashed), $D(1^+)$ (green, short-dashed),
          $D^*(1^+)$ (red, solid) and $D^*(2^+)$ (dark blue, dotted) as a function
          of the parameter $g$. The parameter $g'$ is fixed by the condition $g' = -g$.
          The vertical lines indicate the best fit value $g^*$ for $g$
          and the zero of $\Delta(g, g' = -g)$.}
    \label{splitting}
    \vspace{0.2cm}
\end{figure}

Starting from the heavy quark limit, the splitting within the two doublets is induced by $1/m_c$
effects.  Ignoring their mixing for the moment, the two $1^+$ states will cross when we switch on
the $1/m_c$ terms, leading to the ``level inversion'' observed in the spectrum depicted in
figure~\ref{splitt}.\footnote{%
    In fact, from the current data shown in table~\ref{tab1} only the
    central values indicate a level inversion. Within their uncertainties, the masses for the two
    $1^+$ states may also allow the naively-expected mass hierarchy. Nevertheless, the two states
    are at least very close in mass.
} Switching on the mixing leads to the well known ``level repulsion'', which is shown in
figure~\ref{splitting} for the parameter values obtained from the fit discussed below. For the $1^+$
states this implies a mixing, which is maximal at the crossing point $|\theta| = 45^\circ$. In order
to explain the assumed ``level inversion'', we must have $45^\circ \leq |\theta| \leq 90^\circ$;
i.e., mixing beyond the crossing point.

For a quantitative analysis, we use our simple assumptions and encode
the mixing of the two $1^+$ states in the $2 \times 2$ sub matrix of the Hamiltonian
\begin{equation} \label{2Ham}
{\bf H} = \left[ \begin{array}{cc}   M & a \\ a & M^* \end{array} \right]\,,
\end{equation}
and we define
\begin{align}
    M   & \equiv  M_{1/2} + \frac{1}{4} g - \frac{1}{12} g'\,,  \\
    M^* & \equiv  M_{3/2} - \frac{5}{4} g - \frac{5}{12} g'\,,  \\
\Delta  & \equiv  M^* - M =  M_{3/2} - M_{1/2}  - \frac{3}{2} g - \frac{1}{3} g'\,, \\
      a & \equiv  \frac{\sqrt{2}}{3} g'\,.
\end{align}

For vanishing $g'$ the Hamiltonian becomes diagonal, but in this case the splitting in the
$D^*(J^+)$ doublet is twice as large as the one in the $D(J^+)$.  We note that the splittings within
each of the doublets is equal if $g = -g'$, which we consider as a benchmark point.  This point is
interesting to consider in terms of our simple model: Assuming that all overlap integrals are equal,
we obtain from eq.~(\ref{Blight})
\begin{equation}
    \vec{B} \sim  g  \vec{J} + g' \vec{s} = g ( \vec{J} - \vec{s}) = g \vec{L}\,,
\end{equation}
which would mean that the spectrum is roughly driven by the orbital angular momentum $L$ of the
light degrees of freedom.  As a result, the spectrum is independent of the orientation of the light
quark spin.  However, this could as well be an artifact of our simple assumptions.

To this end, the eigenstates of ${\bf H}$ (i.e., the physical $1^+$ states) are thus linear
combinations of the states defined in the heavy-mass limit
\begin{align}  \label{mixedstates}
| D_L (1^+) \rangle & = \quad\!\cos  \theta  \, | D (1^+) \rangle + \sin  \theta \, | D^*(1^+) \rangle\,, \\
| D_H (1^+) \rangle & = -      \sin  \theta  \, | D (1^+) \rangle + \cos  \theta \, | D^*(1^+) \rangle\,,
\end{align}
where $D_L$ ($D_H$) is the state with the lower (higher) eigenvalue.
The mixing angle $\theta$ satisfies
\begin{equation}
    \tan 2 \theta = \frac{-2 a}{\Delta} = \frac{-4 \sqrt{2} g'}{6 (M_{3/2}-M_{1/2}) - 9g + 2g'}\,,
\end{equation}
and the corresponding eigenvalues are given by
\begin{equation}
    M_{H/L} = \frac{1}{2} \left[ (M+M^*) \pm \sqrt{\Delta^2 + 4 a^2} \right]\,.
\end{equation}

\begin{figure}[t]
    \centering
    \includegraphics[width=0.45\textwidth]{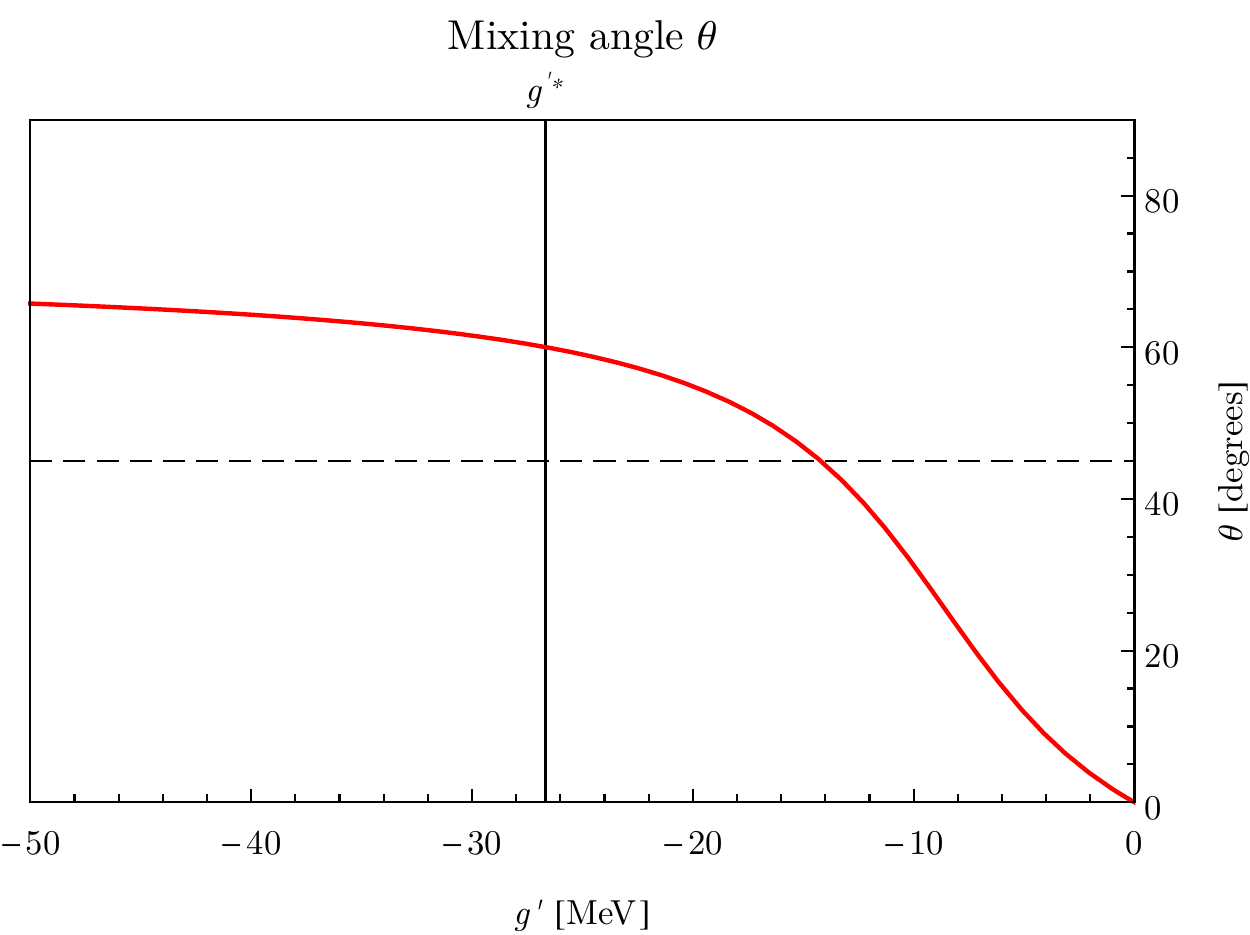}
    \caption{The mixing angle $\theta$ as a function of $g'$ and with $g = -g'$. The vertical
    line indicates the best fit point. The horizontal line indicates the maximum-mixing angle
    of $\theta = 45^\circ$.}
    \label{figcos}
\end{figure}

Note that we have always $M_H > M_L$, reflecting the ``level repulsion'' of the $2\times 2$ system.
The situation of minimal splitting $M_H - M_L$ corresponds to $\Delta = 0$, which yields $|\theta| =
45^\circ$.  This means that the contribution of the $j=3/2$ state in the eigenvector $| D_L (1^+)
\rangle$ starts to become dominant, in accordance to the observed level sequence. This is shown in
figure~\ref{figcos}, where the mixing angle $\theta$ is plotted as a function of $g'$, with the
constraints $g = -g'$.  There, the vertical line corresponds to the best-fit value of $g'$ of the
fit discussed below, where we find a mixing angle of about $60^\circ$.

In previous analyses (e.g.\ in reference \cite{Falk:1995th}), a value for $\theta$ of the order of
ten degrees has been considered. However, this is not in contradiction to our discussion, since an
interchange of the two $1^+$ states corresponds to a replacement $\theta \to 90-\theta$.
Nevertheless, we advocate that the magnitude of $\theta$ is larger than considered in previous
analyses.\\

The experimental results for the masses and widths of the orbitally excited $D$ mesons are presently
unsettled.  The individual results for the masses, for instance, do not agree well with their
respective PDG averages \cite{Agashe:2014kda}.  This is evidenced by the application of scale
factors to the error on the average values. Nevertheless, we attempt to confront our simple model
with the present data.

\begin{table}
    \renewcommand{\arraystretch}{1.2}
    \begin{center}
    \begin{tabular}{|c|c|cc|}
        \hline
        Mass [MeV]         & Pull [$\sigma$]      & \multicolumn{2}{|c|}{Reference}\\
        \hline
        \hline
        \multicolumn{4}{|c|}{$D(0^+)$}\\
        \hline
        $2297   \pm  8   $ & $-13.75$             & \cite{Aubert:2009wg}        & $\times$\\
        $2308   \pm 36   $ & $-2.75$              & \cite{Abe:2003zm}           & $\times$\\
        $2407   \pm 41   $ & $-0.00$              & \cite{Link:2003bd}          & \\
        \hline
        \multicolumn{4}{|c|}{$D_L(1^+)$}\\
        \hline
        $2423.1 \pm  1.8 $ & $+1.24$              & \cite{Abramowicz:2012ys}    & \\
        $2420.1 \pm  0.8 $ & $-0.95$              & \cite{delAmoSanchez:2010vq} & \\
        $2426.0 \pm  3.2 $ & $+1.61$              & \cite{Abe:2004sm}           & \\
        \hline
        \multicolumn{4}{|c|}{$D_H(1^+)$}\\
        \hline
        $2427   \pm 36   $ & $-0.62$              & \cite{Abe:2003zm}           & \\
        $2477   \pm 40   $ & $+0.69$              & \cite{Aubert:2006zb}        & $\dagger$\\
        \hline
        \multicolumn{4}{|c|}{$D^*(2^+)$}\\
        \hline
        $2462.5 \pm  2.7 $ & $+0.01$              & \cite{Abramowicz:2012ys}    & \\
        $2462.2 \pm  0.8 $ & $-0.34$              & \cite{delAmoSanchez:2010vq} & \\
        $2464.5 \pm  2.2 $ & $+0.92$              & \cite{Link:2003bd}          & \\
        \hline
    \end{tabular}
    \end{center}
    \caption{Masses for the various $D^{**}$ measurements. We use up to three of the most
        precise mass measurements, if available. We also list the pull values at the best-fit point eq.
        \ref{eq:results-reduced}.
        Measurements with a $\times$ mark are not included in the fit, see text. Uncertainties of Measurements with
        a $\dagger$ mark have been doubled, due to lack of estimates for the systematic uncertainty.
    }
    \label{tab2}
\end{table}

In a first step, we assume $g' = -g$, thereby reducing the number of model parameters to three:
$M_{1/2}$, $M_{3/2}$, and $g$. This assumption stems from the fact, that the mass splitting of the
doublets is equal for $g = -g'$, as argued above.  We then fit the three parameters to the four PDG
averages of the measurements as listed in table \ref{tab1}. Note, that we impose the additional
constraint $\Delta < 0$, based on the mass hierarchy shown in figure~\ref{splitt}.  This fit has one
degree of freedom (d.o.f.). We reject the fit, since we obtain $\chi^2 = 8.48$ and a p-value of
$0.36\%$, which is smaller than our a-priori threshold of $3\%$. While the masses of the two $1^+$
eigenstates and the $2^+$ state are modelled very well, we find a large pull of slightly less than
$3\sigma$ for the $0^+$ mass.  The best-fit point and the parameter intervals at $68\%$ confidence
level (CL) read
\begin{equation}
\begin{aligned}
    M^*_{1/2} & = (2425 \pm 2) \text{MeV}\,, &
    M^*_{3/2} & = (2451 \pm 5) \text{MeV}\,, \\
    g^*       & = (  23 \pm 6) \text{MeV}\,.
\end{aligned}
\end{equation}

In a second step, we wish to find out which of the individual measurements do not agree well with
our model.  For this purpose, we use the 11 most precise individual measurements, with up to three
measurements per $D^{**}$ state. These measurements also enter the PDG averages, and they are listed
in table \ref{tab2}. Given the larger number of measurements, we can now lift the previous
assumption and fit all four model parameters. We again impose the theory constraint $\Delta < 0$. We
reject this fit as well, since for seven d.o.f.\ we obtain $\chi^2 = 177$, corresponding to a
p-value of less than $10^{-10}$. However, we observe that the $\chi^2$ is driven by two
measurements: the measurements of the $0^+$ mass as carried out by the BaBar and Belle
collaborations, respectively.  It is therefore interesting to repeat this second fit without the
BaBar and Belle measurements of $0^+$ masses, which we do. We find for this new fit the best-fit
point and the $68\%$ CL intervals
\begin{equation}
\label{eq:results-reduced}
\begin{aligned}
    M^*_{1/2} & = (2433 \pm 4) \text{MeV}\,, &
    M^*_{3/2} & = (2450 \pm 5) \text{MeV}\,, \\
    g^*       & = (  26 \pm 5) \text{MeV}\,, &
    g'^*      & = ( -27 \pm21) \text{MeV}\,.
\end{aligned}
\end{equation}
We also compute the goodness of fit for this reduced data set. We find a good fit for five d.o.f.\
and $\chi^2 = 6.86$, which corresponds to a p-value of $0.23$. The individual pull values at the
best-fit point, defined via
\begin{equation}
    \operatorname{pull}_i \equiv \frac{M_i - M_i(M^*_{1/2}, M^*_{3/2}, g^*, g'^*)}{\sigma_i}
\end{equation}
for every measurement $i$, are listed in table \ref{tab2}.  Based on the results of this fit,
eq.~(\ref{eq:results-reduced}), we find approximately $\Delta \simeq -13$ MeV, as well as $g' \simeq
-g < 0$.  This indicates a large mixing, with $\theta = (+59^{+24}_{-14})^{\circ}$ at $68\%$ CL.\\

Our findings can be summarized as follows: Neither the averages, nor the three most precise
measurements of each of the $D^{**}$ masses are fitted well by our simple assumptions.  Removing two
of the individual measurements of the $0^+$ mass from our analysis yields a good fit.  In addition,
most of the measurements are not in good agreement with each other \cite{Agashe:2014kda}. On the
basis of the current experimental data, the situation remains inconclusive.  In our opinion, a
simultaneous determination of the masses and widths of all orbitally excited states needs to be
undertaken before a definite answer to mixing of the $1^+$ states can be given. Within such an
analysis, the mixing effects of the decay widths could be taken into account as well.

\section{Effects on Semileptonic Decays}

\begin{table}[t]
\centering
\renewcommand{\arraystretch}{1.3}
\begin{tabular}{|c||c|c|c|c|}
    \hline
      &$\tau_{1/2}(1)$&$\rho_{1/2}^2$&$\tau_{3/2}(1)$&$\rho_{3/2}^2$\\[0.5ex]
    \hline
    \hline
    GI   \cite{Godfrey:1985xj}              &$0.22$ & $0.83$ & $0.54$ & $1.50$\\
    VD   \cite{Veseli:1996yg}               &$0.13$ & $0.57$ & $0.43$ & $1.39$\\
    CCCN \cite{Cea:1988yd,Colangelo:1990rv} &$0.06$ & $0.73$ & $0.51$ & $1.45$\\
    ISGW \cite{Isgur:1988gb}                &$0.34$ & $1.08$ & $0.59$ & $1.76$\\
    \hline
\end{tabular}
\caption{Results of the fit of eq.~(\ref{tauexp}) to four different model calculations for the two form factors.
    (numbers taken from \cite{Morenas:1997nk}).
}
\label{tb1}
\end{table}

Exclusive semileptonic $b \to c$ decays are governed by heavy quark symmetry for both the bottom and
the charm quark. The decays of $B$ mesons into orbitally excited $D$ mesons have been investigated
in the context of heavy quark symmetry in \cite{Leibovich:1997em}, including also the $1/m_c$
corrections. As stated above and in contrast to \cite{Leibovich:1997em}, we study a scenario where
the mixing is the leading $1/m_c$ effect.

In the infinite mass limit, the hadronic weak transition currents can be described by a single Isgur
Wise function for each multiplet.  Using the spin representations for the states involved
\begin{align}
    \overline{B} (v)
    & = \sqrt{M_B} \gamma_5 \frac{1+\slashed{v}}{2}\,, \\
    D (1^+;v,\epsilon)
    & = \sqrt{M_D} \frac{1+\slashed{v}}{2} \gamma_5 \slashed{\epsilon}\,, \\
    D^*_\mu (1^+;v,\epsilon)
    & = \sqrt{M_{D^*}} \sqrt{\frac{3}{2}} \frac{1+\slashed{v}}{2} \gamma_5\\
    & \quad \nonumber \left[\epsilon_\mu - \frac{1}{3}\slashed{\epsilon} (\gamma_\mu - v_\mu)\right]\,,
\end{align}
one defines for some current defined by a Dirac matrix $\Gamma$  the two Isgur-Wise functions as \cite{Leibovich:1997em}
\begin{multline}
    \langle B (v) |\bar{b}\Gamma c| D(1^+;v',\epsilon) \rangle\\
    = 2 \, \tau_{1/2} (vv') {\rm Tr} \left[ \overline{B}(v) \Gamma  D (1^+;v',\epsilon) \right]\,,
\end{multline}
\begin{multline}
    \langle B (v) |\bar{b}\Gamma c| D^*(1^+;v',\epsilon) \rangle\\
    = \sqrt{3}  \, \tau_{3/2} (vv')  v^\mu {\rm Tr} \left[ \overline{B}(v) \Gamma  D^*_\mu
    (1^+;v',\epsilon) \right]\,.
\end{multline}
In the above the convention for the two Isgur-Wise functions is  the same as in \cite{Leibovich:1997em} and
\cite{Biancofiore:2013ki}.

\begin{widetext}
The mixing induced by $1/m_c$ effects yields for the hadronic currents
\begin{equation}
\label{Mixed}
\begin{aligned}
    \langle B|\bar{b}\Gamma c|D_L\rangle
    & = \quad\! \cos\theta\langle B|\bar{b}\Gamma c|D(1^+) \rangle + \sin\theta\langle B|\bar{b}\Gamma c|D^*(1^+) \rangle\,,\\
    \langle B|\bar{b}\Gamma c|D_H\rangle
    & =        -\sin\theta\langle B|\bar{b}\Gamma c|D(1^+) \rangle + \cos\theta\langle B|\bar{b}\Gamma c|D^* (1^+) \rangle\,,
\end{aligned}
\end{equation}
which can be expressed in terms of the form factors $\tau_{1/2}(\omega)$ and $\tau_{3/2}(\omega) $.
\end{widetext}

The two form factors are parametrized \cite{Morenas:1997nk} via
\begin{equation}  \label{tauexp}
    \tau_j(\omega)=\tau_j(1)\left[\frac{2}{1 + \omega}\right]^{2\rho_j^2},\qquad j=1/2\text{ or }j=3/2
\end{equation}
which is expected to be a reasonable approximation, since the kinematic region turns out to be quite
limited: $1 \le \omega \lesssim 1.3$.

Not much is known about the form factors $\tau_j(\omega)$. However, there are sum rules constraining
these form factors \cite{Uraltsev:2000ce}; in particular we have
\begin{equation}
    \mu_\pi^2  - \mu_G^2  \le 9 \, \epsilon_{1/2} | \tau_{1/2} (1) |^2\,,
\end{equation}
where $\mu_\pi$ and $\mu_G$ are the kinetic energy and chromomagnetic moment parameters, and
$\epsilon_{1/2}$ is the excitation energy of the $j = 1/2$ doublet above the ground state doublet.
Numerically, the left-hand side of this relation is small: $\mu_\pi^2  - \mu_G^2 \ll \mu_\pi^2$.
This motivates the so-called BPS limit, which leads to $ \mu_\pi^2  - \mu_G^2 = 0$. In this limit we
would have $\tau_{1/2} (1) = 0$ indicating that the theoretical expectation is that (in the relevant
kinematic region)
\begin{equation} \label{taus}
\tau_{1/2} (\omega)  \ll \tau_{3/2} (\omega)\,.
\end{equation}
In the combined infinite mass and BPS limits this leads to the expectation that the decays into the
$D (1^+)$ are heavily suppressed compared to the ones into the  $D^* (1^+)$. Current data do not
support this, which constitutes the $1/2-3/2$ puzzle in semileptonic $B$ decays.

Various models and sum rule calculations have been used to obtain more information on the
$\tau_j(\omega)$. In Table~\ref{tb1} we list some of the currently used models in this context; note
that all models reflect the relation eq.~(\ref{taus}) in a more or less pronounced way.

\begin{table}[t]
\begin{center}
\renewcommand{\arraystretch}{1.3}
\begin{tabular}{|l|c|}
\hline
Decay mode                                         & $\mathcal{B}$ (\%)          \\
\hline
\hline
${\mathcal B} (B^- \to D(0^+)    \ell \nu) \times {\mathcal B} (D(0^+)   \to D^+\pi^-)$    & $0.29  \pm 0.05$   \\
${\mathcal B} (B^- \to D_L(1^+)  \ell \nu) \times {\mathcal B} (D_L(1^+) \to D^{*+}\pi^-)$ & $0.29  \pm 0.14$   \\
${\mathcal B} (B^- \to D_H(1^+)  \ell \nu) \times {\mathcal B} (D_H(1^+) \to D^{*+}\pi^-)$ & $0.13  \pm 0.04$   \\
${\mathcal B} (B^- \to D^* (2^+) \ell \nu) \times {\mathcal B} (D^*(2^+) \to D^{*+}\pi^-)$ & $0.078 \pm 0.008$  \\
\hline
\end{tabular}
\end{center}
\caption{HFAG averages for the products of the production branching fractions of $B^- \to D^{**} \ell \nu$ decays
    with the decay branching fractions $D^{**} \to D^{(*)+}\pi^-$~\cite{Amhis:2014hma}.
    The uncertainties arise from the squared sum of the statistical and systematical uncertainties.
}
\label{tb:tau}
\end{table}

\begin{widetext}
The differential decay rates in the infinite mass limit are well known \cite{Leibovich:1997em}.
Including the mixing eq.~(\ref{Mixed}) as the leading $1/m_c$ effect one obtains for $B\to
D_L\ell\bar{\nu}$
\begin{equation}
\label{eq:H}
\begin{aligned}
    \frac{d\Gamma}{d\omega}&=\frac{G_F^2V_{cb}^2M_B^5}{24\pi^3}\,r_L^3(\omega-1)\sqrt{\omega^2-1}\\
    \Big\{&\quad\sin^2\theta(\omega+1)^2[2(\omega-r_L)(1-r_L\omega)-(1+r_L^2-2r_L\omega)]|\tau_{3/2}(\omega)|^2\\
    &+2\cos^2\theta[2r_L(\omega^2-1)+(5\omega-1)(1+r_L^2-2r_L\omega)]|\tau_{1/2}(\omega)|^2\\
    &+2\sqrt{2}\sin\theta\cos\theta(\omega+1)^2[(1+r_L)^2-4r_L\omega]Re(\tau_{1/2}(\omega)\tau^*_{3/2}(\omega))\Big\}
    \,,
\end{aligned}
\end{equation}
while we get for $B\to D_H\ell\bar{\nu}$
\begin{equation}
\label{eq:L}
\begin{aligned}
    \frac{d\Gamma}{d\omega}&=\frac{G_F^2V_{cb}^2M_B^5}{24\pi^3}\,r_H^3(\omega-1)\sqrt{\omega^2-1}\\
    \Big\{&\quad\cos^2\theta(\omega+1)^2[2(\omega-r_H)(1-r_H\omega)-(1+r_H^2-2r_H\omega)]|\tau_{3/2}(\omega)|^2\\
    &+2\sin^2\theta[2r_H(\omega^2-1)+(5\omega-1)(1+r_H^2-2r_H\omega)]|\tau_{1/2}(\omega)|^2\\
    &-2\sqrt{2}\sin\theta\cos\theta(\omega+1)^2[(1+r_H)^2-4r_H\omega]Re(\tau_{1/2}(\omega)\tau^*_{3/2}(\omega))\Big\}\,.
\end{aligned}
\end{equation}
The above rates now depend on both of the form factors $\tau_j (\omega)$, as well as on
the mixing angle $\theta$. By inserting the expression eq.~(\ref{tauexp}) with the values given in
table~\ref{tb1} and using the nominal fit result $\cos\theta\simeq 0.51$, we calculate
the invidual branching ratios shown in table~\ref{tab:Pred}.
\end{widetext}

\begin{table*}[t]
\centering
\renewcommand{\arraystretch}{1.3}
\newcommand{\sci}[1]{\cdot 10^{#1}}
\begin{tabular}{|c||c|c|c|c|}
    \hline
    Channel &GI&VD&CCCN&ISGW\\[0.5ex]
    \hline
    \multicolumn{5}{|c|}{$m_c \to \infty$}\\
    \hline
    $\mathcal{B}(B^-\to D (0^+)\ell\bar{\nu})$  &$4.7\sci{-4}$ & $1.8\sci{-4}$ & $3.7\sci{-5}$ & $1.0\sci{-3}$\\
    $\mathcal{B}(B^-\to D (1^+)\ell\bar{\nu})$  &$6.4\sci{-4}$ & $2.5\sci{-4}$ & $4.9\sci{-5}$ & $1.4\sci{-3}$\\
    $\mathcal{B}(B^-\to D^*(1^+)\ell\bar{\nu})$ &$4.4\sci{-3}$ & $2.9\sci{-3}$ & $4.0\sci{-3}$ & $4.7\sci{-3}$\\
    $\mathcal{B}(B^-\to D^*(2^+)\ell\bar{\nu})$ &$7.4\sci{-3}$ & $4.9\sci{-3}$ & $6.7\sci{-3}$ & $8.0\sci{-3}$\\
    \hline
    $\mathcal{B}(B^-\to D^{**}\ell\bar{\nu})$   &$1.3\%$       & $0.82\%$      & $1.1\%$       & $1.5\%$\\
    \hline
    \phantom{\Bigg|}
    $\dfrac{\mathcal{B}(B^-\to D^*(1^+)\ell\bar{\nu})}{\mathcal{B}(B^-\to D(1^+)\ell\bar{\nu})}$
                                                &$6.9$         & $11$          & $80$          & $3.4$\\
    \hline
    \multicolumn{5}{|c|}{$m_c$ finite}\\
    \hline
    $\mathcal{B}(B^-\to D_L\ell\bar{\nu})$      &$3.0\sci{-3}$ & $2.1\sci{-3}$ & $3.0\sci{-3}$ & $3.2\sci{-3}$\\
    $\mathcal{B}(B^-\to D_H\ell\bar{\nu})$      &$2.3\sci{-3}$ & $1.3\sci{-3}$ & $1.3\sci{-3}$ & $3.1\sci{-3}$\\
    \hline
    \phantom{\Bigg|}
    $\dfrac{\mathcal{B}(B^-\to D_L\ell\bar{\nu})}{\mathcal{B}(B^-\to D_H\ell\bar{\nu})}$
                                                &$1.3$         & $1.6$         & $2.3$         & $1.0$\\
    \hline
\end{tabular}
\caption{Predictions for the branching fractions for the channels $B\to D^{**}\ell\bar{\nu}$
    within the various models of table~\ref{tb1}, both with and without mixing effects among the
    $1^+$ states.  Here $\mathcal{B}(B^-\to D^{**}\ell\bar{\nu})$ represents the sum of four
    branching fractions to the $0^+$, the two $1^+$ and the $2^+$ states.  Note, that for $m_c\to
    \infty$, we use the doublet masses $M_{1/2}$ and $M_{3/2}$ as inputs, while for the finite $m_c$
    results, we use the experimentally determined masses.
}
\label{tab:Pred}
\end{table*}

The first six rows of table~\ref{tab:Pred} are the values obtained in the infinite mass limit, where
we use the experimental results for the lifetimes. The last three rows are obtained for finite
$m_c$, where we consider mixing effects among the $1^+$ states as the leading $1/m_c$ effect.

Given the large uncertainties on the experimental inputs, we wish to emphasize that our analysis is
only meant as a qualitative study; i.e., to answer the question: Can mixing between the two $1^+$
states (at least partially) explain the $1/2$ -- $3/2$ puzzle?  As a consequence, we abstain from
providing uncertainty estimates on the quantities in table~\ref{tab:Pred}.  Clearly we find a strong
impact of the mixing, which roughly swaps the roles of the two $1^+$ states.  We find that the
inclusion of $1/m_c$ mixing effects redistributes the relative weights of the invidivual decay
channels within the decays $B^-\to D^{**}\ell\bar\nu$.

Nevertheless, our results can be confronted with the experimental data shown in Table~\ref{tb:tau}.
Unfortunately, there are no experimental results yet available on the absolute branching fractions,
since the branching fractions for the subsequent strong decays $D^{**}\to D^{(*)+}\pi^-$ have not
yet been measured. However, assuming that these subsequent decays have roughly the same branching
fractions, the measured ratio of the two decays into $1^+$ states is approximately
\begin{equation}
    \frac{\mathcal{B}(B^- \to D_L(1^+)  \ell \nu) }{\mathcal{B}(B^- \to D_H(1^+)  \ell \nu) } \approx 2.2\,.
\end{equation}
The estimates without mixing effects (table~\ref{tab:Pred}, row 6) clearly deviate from the measured
ratio. On the other hand, our estimates with mixing effects taken into account
(table~\ref{tab:Pred}, row 9) are in reasonable agreement with the measurements.

\section{Effects on the Widths of the orbitally excited states}

In \cite{Isgur:1991wq} it has been discussed that in the infinite mass limit one has the relations
\begin{align}
    {\cal A} (D^* (2^+) \to D \pi)   & \propto \sqrt{\frac{2}{5}} a_D \\
    {\cal A} (D^* (2^+) \to D^* \pi) & \propto \sqrt{\frac{3}{5}} a_D \\
    {\cal A} (D^* (1^+) \to D^* \pi) & \propto a_D                    \\
    {\cal A} (D^* (1^+) \to D \pi)   & = 0                            \\
    {\cal A} (D(1^+) \to D^* \pi)    & \propto a_S                    \\
    {\cal A} (D(1^+) \to D \pi)      & = 0                            \\
    {\cal A} (D (0^+) \to D \pi)     & \propto a_S                    \\
    {\cal A} (D(0^+) \to D^* \pi)    & = 0
\end{align}
where $a_D$ and $a_S$ are the amplitudes for the $D$-wave and the $S$-wave decays of the $D^{**}$ states.

We assume that these modes dominate the total widths, such that
\begin{equation}
\Gamma_{\rm tot} (D^{**}) = \Gamma (D^{**} \to D \pi) + \Gamma ( D^{**} \to D^* \pi)  \, ,
\end{equation}
so we obtain the predictions in the heavy quark limit
\begin{equation}
\begin{aligned}
    \Gamma_{\rm tot} (D^*(2^+)) & = |a_D|^2 = \Gamma_{\rm tot} (D^*(1^+))\,, \\
    \Gamma_{\rm tot} (D(0^+))   & = |a_S|^2 = \Gamma_{\rm tot} (D(1^+))\,,
\end{aligned}
\end{equation}
where we ignore small phase space differences of the order of twenty percent.  Since we expect the
$D$-wave amplitude to be suppressed relative to the $S$-wave amplitude by the usual angular momentum
factors, we arrive at the well known conclusion that the $j = 1/2$-doublet states are broader than
the states of the $j = 3/2$ doublet.

Including the mixing induced by the $1/m_c$ terms yields the relations
\begin{align}
    {\cal A} (D_L(1^+) \to D\pi)    & = 0 \, = {\cal A} (D_H (1^+) \to D\pi)\,,\\
    {\cal A} (D_L (1^+) \to D^*\pi) & = a_S \, \cos \theta  + a_D \, \sin \theta\,,\\
    {\cal A} (D_H (1^+) \to D^*\pi) & = a_D \, \cos \theta   - a_S \, \sin \theta\,.
\end{align}
Within the differential decay width, interference terms between the $S$ and $D$ wave amplitudes
arise. However, they drop out after integrating over the $D^*\pi$ helicity angle.
Consequently we have
\begin{multline}
    \label{MixedWidths1}
    \Gamma_{\rm tot} (D_L) = \Gamma_L\\
        \sim \Gamma(D_L \to D^* \pi) = |a_S|^2 \, \cos^2 \theta + |a_D|^2 \, \sin^2 \theta\,,
\end{multline}
and
\begin{multline}
    \label{MixedWidths2}
    \Gamma_{\rm tot} (D_H) = \Gamma_H\\
        \sim \Gamma(D_H \to D^* \pi) = |a_D|^2 \, \cos^2 \theta + |a_S|^2 \, \sin^2 \theta\,,
\end{multline}
where we again ignore small phase space differences.

The experimental situation on strong decays of the $D^{**}$ states is shown in table~\ref{tab1} and is not yet
conclusive. Nevertheless, we can get a qualitative picture by assuming that we can extract $|a_S|$
and $|a_D|$ from the widths of the $0^+$ and the $2^+$ states, respectively, and insert these into
eq.~(\ref{MixedWidths1}). From this we obtain
\begin{equation*}
    \Gamma_L \sim  240 \, {\rm MeV} \,, \qquad
    \Gamma_H \sim 80 \, {\rm MeV}\,,
\end{equation*}
where we again do not consider the experimental uncertainties, since we only aim at the qualitative
picture.

Thus the observed pattern is not in contradiction to a large mixing of the $1^+$ states, although
the observed factor of about two between the widths of the two narrow states still remains
unexplained.

\section{Conclusion}

We have discussed a scenario where the mixing of the two $1^+$ states of the first orbitally excited
charmed mesons is assumed to be large. Such a large mixing can still be accommodated with the data.
It is supported by very simple arguments on the physics origin of such a mixing and some
assumptions on the size of some matrix elements.  However, the input for any estimate is the data on
the masses and the widths. The data are not yet conclusive, at least not for the broader one of the
$1^+$ states.

If such a large mixing is indeed present, it will also have consequences for the view on
spectroscopy from the perspective of the heavy quark  limit. At least for the charm mesons this
means that the spin-symmetry doublets will have a significant mixing for the states with the same
quantum numbers, which is induced by interactions that are formally of subleading order but are
numerically significant.

For the semileptonic decays of $B$ mesons such a mixing would soften the 1/2 -- 3/2 puzzle at least
for the the $1^+$ states, since a mixing with a significant angle will reduce the difference of the
two rates. Nevertheless, in order to pin down if the mixing is really the solution to this puzzle,
more data on the semileptonic decays as well as on the masses and the widths of the orbitally
excited $D$ meson states will be required.

\acknowledgments

This work was supported by BMBF and the DFG Research Unit FOR 1873. FS acknowledges support by a
Nikolai Uraltsev Fellowship of Siegen University.  We thank Sascha Turczyk for valuable discussions.

\appendix

\section{Spin Wave Functions}
In this appendix we give the explicit formulae for the coupling of the angular momentum $L = 1$ and the light quark spin $s = 1/2$.
The spin wave functions for the case $j = 1/2$ read
\begin{align}
    \label{j1}
    | j \!=\!1/2, +1/2 \rangle
    & = \sqrt{\frac{2}{3}} | 1 \rangle |\!-\!1/2 \rangle_l  -  \sqrt{\frac{1}{3}} |  0 \rangle |1/2 \rangle_l \\
    | j \!=\!1/2, -1/2 \rangle
    & = \sqrt{\frac{1}{3}} | 0 \rangle |\!-\!1/2 \rangle_l  -  \sqrt{\frac{2}{3}} |\!-\!1 \rangle |1/2 \rangle_l
\end{align}
and for $j  = 3/2$ we get
\begin{align}
    | j = 3/2, +\!3/2 \rangle & =  |\!+\!1 \rangle |\!+\!1/2 \rangle_l  \vphantom{\sqrt{\frac{1}{3}} }  \\
    | j = 3/2, +\!1/2 \rangle & =  \sqrt{\frac{1}{3}} | 1 \rangle |\!-\!1/2 \rangle_l  +  \sqrt{\frac{2}{3}} | 0 \rangle | 1/2 \rangle_l \\
    | j = 3/2, -\!1/2 \rangle & =  \sqrt{\frac{2}{3}} | 0 \rangle |\!-\!1/2 \rangle_l  +
    \sqrt{\frac{1}{3}} |\!-\!\!1\rangle | 1/2 \rangle_l \\
    | j = 3/2, -\!3/2 \rangle & =  |\!-\!1 \rangle |\!-\!1/2 \rangle_l  \vphantom{\sqrt{\frac{1}{3}} }  \label{jn}
\end{align}
where the first ket vector is for the angular momentum, while $| \cdot \rangle_l$ denotes the spin of the light quark.

The above states have to be combined with the heavy quark spin in order to obtain the spin wave
functions of the $D^{**}$ mesons. Since we are interested in the mixing of the $1^+$ states, we
concentrate on these states and obtain
\begin{widetext}
\begin{align}
| D(1^+), M=1 \rangle  &= | j  = 1/2, 1/2 \rangle | 1/2 \rangle_H \\
| D(1^+), M=0 \rangle  &= \sqrt{\frac{1}{2}} \left(  | j = 1/2, 1/2 \rangle | - 1/2 \rangle_H
  +   | j  = 1/2, - 1/2 \rangle | 1/2 \rangle_H   \vphantom{\sqrt{\frac{1}{3}} }  \right) \\
| D(1^+), M=-1 \rangle &= | j  = 1/2, - 1/2 \rangle | - 1/2 \rangle_H
\end{align}
and
\begin{align}
| D^*(1^+), M=1 \rangle  &= \sqrt{\frac{3}{4}}  | j = 3/2, 3/2 \rangle | -1/2 \rangle_H
     -  \sqrt{\frac{1}{4}}  | j  = 3/2, 1/2 \rangle | 1/2 \rangle_H \\
| D^*(1^+), M=0 \rangle  &= \sqrt{\frac{1}{2}}  | j  = 3/2, 1/2 \rangle | -1/2 \rangle_H
     -  \sqrt{\frac{1}{2}}  | j  = 3/2, -1/2 \rangle | 1/2 \rangle_H \\
| D^*(1^+), M=-1 \rangle &= \sqrt{\frac{1}{4}}  | j = 3/2, - 1/2 \rangle | -1/2 \rangle_H
     -  \sqrt{\frac{3}{4}}  | j  = 3/2, -3/2 \rangle | 1/2 \rangle_H
\end{align}
\end{widetext}
Combining this with eqs.~(\ref{j1}-\ref{jn}) yields the wave functions which can be used to discuss the mixing.

The spin-spin coupling can be computed using
\begin{equation}
    (\vec{s} \cdot \vec{\sigma})  = \frac{1}{2} (s_+ \sigma_- + s_- \sigma_+) + s_3 \sigma_3
\end{equation}
which yields for any $M$ the result
\begin{align}
    (\vec{s} \cdot \vec{\sigma}) | D(1^+) \rangle   & = - \frac{1}{12} | D(1^+) \rangle +
    \frac{\sqrt{2}}{3} | D^*(1^+) \rangle\,, \\
    (\vec{s} \cdot \vec{\sigma}) | D^*(1^+) \rangle & = - \frac{5}{12} | D^*(1^+) \rangle +
    \frac{\sqrt{2}}{3}  | D(1^+) \rangle\,.
\end{align}

\bibliographystyle{apsrev4-1}
\bibliography{references.bib}

\end{document}